# Climate Implications of Diffusion-based Generative Visual AI Systems and their Mass Adoption

**Vanessa Utz and Steve DiPaola**

School of Interactive Arts and Technology (SIAT)
Simon Fraser University
Burnaby, BC V5A 1S6  Canada
vutz@sfu.ca    sdipaola@sfu.ca

## Abstract

Climate implications of rapidly developing digital technologies, such as blockchains and the associated crypto mining and NFT minting, have been well documented and their massive GPU energy use has been identified as a cause for concern. However, we postulate that due to their more mainstream consumer appeal, the GPU use of text-prompt based diffusion AI art systems also requires thoughtful considerations. Given the recent explosion in the number of highly sophisticated generative art systems and their rapid adoption by consumers and creative professionals, the impact of these systems on the climate needs to be carefully considered. In this work, we report on the growth of diffusion-based visual AI systems, their patterns of use, growth and the implications on the climate. Our estimates show that the mass adoption of these tools potentially contributes considerably to global energy consumption. We end this paper with our thoughts on solutions and future areas of inquiry as well as associated difficulties, including the lack of publicly available data.

## Introduction

Many of today's rapidly developing digital technologies are being critically discussed due to their potential climate impact. For instance, blockchain-based technologies and the associated crypto-mining and NFT (Non-Fungible Tokens) minting, have frequently come under fire for their high energy usage (deVries 2019). While AI has often been proposed as a solution to the problems digital technologies create for our environment, interest has now also shifted to machine learning (ML) and its environmental implications. However, so far most of the work in this area has focused on the training phase of large AI systems. Our position is that more effort needs to be put into the study of the environmental impact that AI systems have during their usage (referred to technically as inference), particularly given the recent explosion in commercially available AI art systems that have been shown to have large scale consumer appeal.

While the environmental consequences that are associated with the energy consumption required to power AI tools, are not unique to the creative AI (cAI) field (including Computational Creativity), we want to raise awareness of this issue among cAI researchers and encourage further scientific investigations. It is also our hope to illustrate that this issue can be positioned as a user education concern.

We begin this position paper by providing some foundational background information into the discourse surrounding technology and its climate impact, both good and bad. We then discuss the current state of research into the energy consumption of ML systems. We also provide an overview of popular AI systems for visual art creation and provide data on their scale. To support our position regarding the need for more research into the environmental impact of cAI systems, we then proceed to demonstrate a preliminary analysis of the energy consumption of generative visual AI systems and compare the data to the climate impact of other digital technologies. We end this paper by drawing a connection to contemporary discussions on waste and overconsumption in the digital space. We also highlight future research avenues such as inquiries into the psychological principles that lead to the prolonged use of these tools.

## Technology and Climate

We begin this topic with a brief introduction to the discourse on technology and climate. For this purpose, we will begin talking about technology and climate in general terms, before focusing on our main concern around energy consumption and carbon emissions.

Much of the conversation that views technology in a critical light appears to be focused on *electronic waste* (or *e-waste*). E-waste refers to the waste associated with the discarding of electronic devices that have reached their end of life (EOL) and has long been a focal point in discussions on pollution and the climate, with the Basel Convention labeling e-waste as hazardous over a decade ago (Widmer, Oswald-Krapf, Sinha-Khetriwal, Schnellmann and Böni 2005).

The main issues surrounding this debate involve the planned obsolescence of consumer electronics (Bisshop, Hendlin and Jaspers 2022), the low rate of recycled electronic devices/components (Perkins, Drisse, Nxele and Sly 2014), greenhouse gas (GHG) emissions (Singh and Ogunseitan 2022) and the pollutants contained within the devices, which have been shown to have detrimental health impacts on individuals exposed during the recycling/disposing process (Chen, Dietrich, Huo and Ho 2011).

### Energy Consumption

Apart from the e–waste that is produced through modern day electronics, another concern involving technology and the climate, centers on the energy consumption of electronic devices while they are in use. Particularly with the rise in Blockchain technology much of the contemporary discourse on technology and climate has shifted to the immense energy consumption that is associated with these new technologies. Energy consumption is of concern here as increased energy consumption is linked to increases in GHG emissions (Luccioni and Hernandez-Garcia 2023; Schwartz, Dodge, Smith and Etzioni 2019).

### Blockchain and the Internet of Things

Blockchain technology and the associated mining of cryptocurrencies and minting of NFTs has made headlines across the world over the last few years. In 2018 alone, the Bitcoin mining network was estimated to have consumed between 40 and 62.3 TWh of energy (which has been compared to the electricity consumption of major European countries like Hungary and Switzerland) (deVries 2019). According to the Bitcoin Energy Consumption Index (2023), in the first half of 2022, energy consumption of the Bitcoin network peaked at around 204.5 TWh per year. This peak lasted for approximately the first 6 months of 2022, before sharply declining to 91.31 TWh per year as of February 2023.

Alarm bells have also been going off in the area of the Internet of Things (IoT) and technological proliferation within our homes and environments (smart homes turning into smart cities). The need for investigations into the energy consumption of these systems has been made clear (Moutaib, Fattah and Farhaoui 2020).

### Artificial Intelligence, the Savior?

In much of the literature discussed above, AI is often hailed as the solution to all our problems (for a brief literature review see: Dwivedi et al. 2022). Discussions involve incorporating AI into evermore technologies, such as Blockchain-based technologies, smart tech and data management systems, to increase efficiency and lower energy consumption and carbon emissions. Additionally, AI models are continuously proposed to aid in the mitigation of the impact of climate change. For instance, AI-based solutions have been suggested for aiding the food and agriculture sector (Ayed and Hanana 2021) and aiding with climate modeling (Huntingford, Jeffers, Bonsall, Christensen, Lees and Yang 2019).

Although these proposals are commendable, the impact of AI systems on the climate is rarely discussed in this type of literature. Even when these impacts are mentioned, they oftentimes are quickly glossed over instead of outlined thoughtfully and in much-needed detail (for example: PwC & Microsoft report on "How AI can enable a Sustainable Future" by Jobba and Herweijer n.d.).

### Machine Learning and Energy Consumption

Although missing from much of the climate change focused AI literature, there is work on the environmental implications of ML that has been picking up momentum over the last few years. Energy consumption is a big focus here due to its relation to increased carbon emissions. Research in this area has focused mainly on the training of ML models. Here we will provide a quick overview of some of the work that has been taking place.

Strubell, Ganesh and McCallum (2019) took a closer look at deep learning models for NLP (Natural Learning Processing) and their financial and environmental costs. They posit that as models are becoming increasingly complex, more computational power is required leading to the increased use of powerful GPUs (Graphical Processing Units) and TPUs (Tensor Processing Units). Their study shows that training a model on a GPU has a similar amount of carbon emission as a trans-America flight. Lacoste, Luccioni, Schmidt and Dandres (2019) presented their Machine Learning Emissions Calculator with the goal that it would prove to be a useful tool for the ML community to track and estimate the carbon emissions during the training phase of ML models. More recently, Luccioni and Hernandez-Garcia (2023) published a survey of 95 ML models used in NLP and computer vision tasks. The survey contains data on energy consumption, CO2 emissions and how emissions volume relate to model performance. The results showed significant carbon emissions associated with the models that were reviewed and the authors call for a better understanding of the environmental impact of ML models within the community of researchers and developers.

## The Mass Adoption of Creative AI

In the last few years, major developments have taken place in the cAI space, particularly with the rapid improvements that were seen in diffusion models. Not only are these models being used and tested among ML researchers and developers, but they have also shown to have mainstream consumer appeal and a large number of free as well as paid services and tools have now been developed and made available to the public. Some of the larger systems include DallE-2 (developed by OpenAI), Midjourney, Stable Diffusion (developed by Stability AI) and Artbreeder. It has been estimated that these four services alone produce over 20 million images per day (Note: this number is frequently cited online, however we were unable to independently corroborate this number) (Kelly 2022; Pennington 2022). The consumer appeal of these systems is also easily demonstrated by taking a closer look at the popularity of related mobile applications. For instance, when Google announced their Google Play's Best 2022 awards in November of said year, the best overall app was awarded to Dream by WOMBO (Lim 2022), a diffusion AI art generator that was released on November 16, 2021 (Wombo.ai 2021) and which as of late February 2023 has over 10 million downloads in the Google Play store. Although on WOMBO's website, the company indicates that globally the app has been downloaded over 100 million times, with a total image output of over 750 million (Wombo Inc. n.d.) Shortly afterwards, Lensa AI by Prisma Labs Inc., which was first released in 2018, exploded in popularity in December of 2022, when it reached over 12 million global downloads (Perez 2022), with 5.8 million downloads occurring in the first week of December alone (Ceci 2023a). The sharp increase in user numbers has been associated with the release of the app's new feature "Magic Avatars", which turns portraits into stylized imagery using a diffusion model. During the first weeks of December 2022, users spent approximately USD 9.25 million on the app's subscription and premium features according to data published by Statista (Ceci 2023b).

We will spend the remainder of this section introducing the most influential systems and highlighting their explosive growth over the last year. The purpose of this is to demonstrate the enormous computing power that is required to meet consumer demand. See Table 1 for a summary of user and daily image output numbers.

| System/Platform | Total Users (in million) | Daily Image Output (in million) |
|---|---|---|
| **Dall-E2** | 3 | 4 |
| **Midjourney** | 12[1] | Unknown[3] |
| **Stable Diffusion** | 10 | Unknown[3] |
| **DreamStudio** | 1.5 | 3[4] |
| **Dream** | 10-100[2] | 1.6[4] |
| **LensaAI** | 12[2] | Unknown[3] |

**Table 1 - Summary of user numbers and daily image output of a selection of the most popular generative AI art systems.**
[1]Number estimated based on members on official server (no numbers available for private servers), [2]Number estimated based on app downloads (might not be reflective of daily active users), [3]No estimates available based on limited data, [4]Number estimated based total image output since release date divided by days the system has been available.

## Dall-E & Dall-E2

Dall-E was first released in January 2021, with the follow-up version Dall-E2 released in April 2022. The system was developed by OpenAI. The exact training data set has not been released by OpenAI. Dall-E2 is able to produce text-to-image and image-to-image output and is able to modify existing images (i.e. "inpainting"). Initially, the system was available through an invite-only service but has since been made available to a broader audience. According to OpenAI, as of November 2022, around 3 million people were using Dall-E2 to generate more than 4 million images per day (OpenAI 2022). OpenAI has also recently released an API (Application Programming Interface) which now enables developers to integrate Dall-E2 into their own applications, making the system even more widely available.

## Midjourney

Developed by a research lab with the same name, this system was first released in April 2022. The newest version was released to the public in November 2022. Like OpenAI, Midjourney has not made their training data public. Midjourney operates as a bot through the Discord platform (an online communication platform which is divided into smaller communities, or so-called "servers", and allows text, voice and video chat). While the model as not been released to the public and the codebase and architecture are therefore unknown, according to StabilityAI's CEO Emad Mostaque, Midjourney has been leaning on Stable Diffusion since its beta lease (Mostaque 2022). As of February 2023, we have confirmed that the official Midjourney Discord server has over 12 million paying members (standard subscriptions currently start at USD30 per month). Members on the server are able to utilize the bot by providing text prompts to trigger image generation. It is important to note that on August 2nd, 2022, Midjourney announced on Twitter that the bot could be added to private servers and that users no longer had to use the official Midjourney server to generate images. We reached out to Midjourney but were unable to confirm how many servers the bot is currently operating.

## Stable Diffusion (& DreamStudio)

Stable Diffusion uses a Latent Diffusion Model (LDM). It was trained on 2.3 billion images, contained within three datasets provided by LAION (Large-scale Artificial Intelligence Open Network): LAION-2B-EN, LAION-High-Resolution and LAION-Aesthetics 2v 5+. The initial release of Stable Diffusion took place in August 2022, with the stable release following in December 2022. The system was developed by StabilityAI. During an interview with Bloomberg in October 2022, StabilityAI's CEO confirmed that Stable Diffusion had over 10 million daily users and that their paid service DreamStudio has around 1.5 million active users (Fatunde and Tse 2022). Users on DreamStudio had generated a total of over 170 million images between launch and October 17th, 2022 (a time frame of 56 days) (StabilityAI 2022). StabilityAI has also noted that over 200k developers had downloaded their model (StabilityAI 2022). It is important to note that Stable Diffusion released their code and model to the public, therefore individuals are able to run the system locally on their own machines. Since the model is freely available and only requires under 10 GB of VRAM (video RAM) on widely available consumer GPUs (StabilityAI n.d.), it is difficult to estimate the true number of daily users. The open-source nature of Stable Diffusion has also caused it to be very commonly used as the system that powers many of the mobile AI art apps, such as Lensa AI (Hatmaker 2022), further increasing the difficulty associated with estimating its true reach and daily users.

## The Energy Consumption of Generate AI Art Systems

Based on the numbers of daily users and daily output generation outlined in the previous section, we would like to provide some preliminary calculations into the energy consumption of these systems. It is important to point out that these numbers are at best a vast underestimation of how much energy is actually consumed, since we do not have access to the precise data on the usage of these systems.

We have decided to focus our initial calculations on Stable Diffusion, since proprietary information on, for instance, which data centers are used by Midjourney, introduces additional variables and unknowns into these calculations. We aim to provide a simple initial exploration into this topic.

### Assumption 1: Hardware

According to sources, Stable Diffusion only natively supported NVIDIA RTX GPUs as of December 2022 (although it can be run in limited ways using other GPUs and CPUs) (Lewis 2022). However, according to StabilityAI's FAQ as of February 2023, most NVIDIA GPUs with 6GB or more, and high-end AMD GPUs are supported (n.d.). Additionally, NVIDIA RTX GPUs (particularly the RTX 3090) outperform most other commercially available GPUs in benchmark testing involving Stable Diffusion (Walton 2023). We therefore assume that this hardware is a reasonable scenario for energy consumption calculations. According to NVIDIA the Total Graphics Power (TGP) of the RTX 3090 is 350W, representing peak power draw (Burnes 2022; Cervenka 2022). We independently confirmed this by generating images using Stable Diffusion on a RTX 3090. Energy draw peaked at 350W when we generated a 1024x1024 image using the default 50 steps.

### Assumption 2: Duration of Use

Regarding the duration of use of their hardware, we assume that the average user runs Stable Diffusion requiring peak power draw for approximately 1.5 hours per day. This assumption is based on a survey which we created to collect preliminary quantitative and qualitative data on the typical use of these systems. The survey was posted in seven AI art communities on Facebook and was live for 6 days, during which we collected 42 responses. The survey consisted of 5 multiple choice questions on the motivation behind their typical art creation, purpose of the final output/artwork, the type of system/tool used, their estimated total weekly image output, and their estimated average iteration per final artwork. The survey ended with one open-ended question asking participants to elaborate on their post-processing procedures, reuse and storing of images or any other information they deemed relevant to their creation process. On the question regarding their average weekly image generation, participants most frequently responded that they create over 1000 images per week. Unfortunately, we were unable to collect more specific data on the average output since we frankly did not expect such a high number of image generations for average users and limited our question to a maximum of 1000 images per week. However, during the open-ended section of the survey, a large portion of our respondents indicated they produce hundreds (sometimes thousands) of images in a single day using automated scripts. We therefore assume an average output of approximately 2000 images per week (or approximately 285 images per day). With this broad estimation we are trying to accommodate casual users as well as power-users. These calculations should be updated once better data becomes available. Based on our own testing on a RTX 3090, creating a 1024x1024 image using the default 50 steps, results in a generation time of 20 seconds. Based on this data, we assume that a user can generate up to 180 images per hour on a RTX 3090. To generate the target daily output of 285 images, a user needs to run the system at peak power draw for approximately 95 minutes per day.

## Energy Consumption Calculations

**Total Yearly Energy Consumption**

Daily energy use per user:

$$350 \text{ W} \times 1.5 \text{ h} = 525 \text{ Wh} = 0.525 \text{ kWh} \quad (1)$$

Daily energy use for 10 million users:

$$0.525 \text{ kWh} \times 10{,}000{,}000 = 5{,}250{,}000 \text{ kWh} \quad (2)$$

Yearly energy use for 10 million users:

$$365 \times 5{,}250{,}000 \text{ kWh} = 1{,}916{,}250{,}000 \text{ kWh} \quad (3)$$
$$= 1.92 \text{ TWh}$$

Based on our assumption that the 10 million users of Stable Diffusion (as confirmed by StabilityAI) run the system for approximately 1.5 hours per day on a RTX 3090, will lead to a yearly energy consumption of approximately 1.92 TWh. This level of energy consumption is similar to the total electricity consumption of the West African nation Mauritania in 2021, which has been estimated to be 1.9 TWh according to the US Energy Information Administration (eia.gov n.d.).

**Energy Consumption per Image**

Number of images generated per hour:

$$3600 \text{ s per hour} \div 20 \text{ s required per images} = 180 \text{ images} \quad (4)$$

Energy use per image:

$$350 \text{ Wh} \div 180 \text{ images} = 1.94 \text{ Wh} \quad (5)$$

**Extrapolation to Other Systems**

In the following section, we are aiming to extrapolate the above data to a larger set of popular generative AI art systems that are either using Stable Diffusion code as their base (such as LensaAI) or are using similar diffusion-based technology. We estimate that the five popular systems Stable Diffusion (including DreamStudio), Midjourney, DallE-2, LensaAI, and Dream have approximately 48.5 million users (see Table 1 for a summary of user data).

Daily energy use per user:

$$350 \text{ W} \times 1.5 \text{ h} = 525 \text{ Wh} = 0.525 \text{ kWh} \quad (6)$$

Daily energy use for 48.5 million users:

$$0.525 \text{ kWh} \times 48{,}500{,}000 = 25{,}462{,}500 \text{ kWh} \quad (7)$$

Yearly energy use for 48.5 million users:

$$365 \times 25{,}462{,}500 \text{ kWh} = 9{,}293{,}812{,}500 \text{ kWh} = 9.29 \text{ TWh} \quad (8)$$

To put this estimation into perspective, based on numbers published by the US Energy Information Administration (eia.gov n.d.), the total electricity consumption of Kenya in 2021 was 9.1 TWh.

There are some obvious limitations with these numbers that need to be addressed. 1) Total users/downloads versus daily active users: the current calculations are based mainly on the publicly available data regarding app downloads, server members etc. This obviously differs from the actual number of daily users, since not every person who signs up for a subscription, downloads an app or installs a model on their device, is using the system regularly. For most of these systems (Stable Diffusion being the exception), we are unable to determine how many individuals use them on a daily basis. Making this data available to researchers and the general public would be a great step to increase transparency in this space. 2) Home computers versus cloud computing: our calculations are based on the peak energy draw of a commercially available GPU (RTX 3090) that has been shown to outperform many other processing units in benchmark testing. Apart from the fact that not all users own this specific GPU, we have no insight into how services like Midjourney generate their output. While we reached out to Midjourney, we were unable to receive a clear answer.

## Comparison to Other Digital Technologies

So how does this compare to other digital technologies, such as blockchain related technologies and the training of AI systems, which are already being discussed as potentially being harmful due to their immense energy consumption?

According to the Bitcoin Energy Consumption Index (2023), current energy consumption sits at around 91.31 TWh per year. The energy consumption during the minting of an NFT (on the Ethereum blockchain) has been estimated to be approximately 142 kWh (Kshetri and Voas 2022) (this estimation was based on the energy use of the Ethereum blockchain before *The Merge*, a term referring to the moment on September 15, 2022 when the Ethereum blockchain moved from proof of work to proof of stake; a decision that lowered the blockchain's total energy demand by as much as 99.9996% (deVries 2022)).

Based on these numbers, it becomes evident that our estimation of the energy consumption of running Stable Diffusion on a home computer is vastly smaller than the total energy consumption associated with blockchain-based technology. However, expanding our estimation to include other systems, the total energy consumption becomes considerably more concerning. We would also like to 1) reiterate that this estimation is most likely a vast underestimation of the actual energy consumption involved in the entire AI art domain and 2) highlight how important an early call to action is, considering the potential and vast application spaces of these new tools. In our survey, 75.61% of participants responded that they would move from image generation to video generation once the tools become more widely available and easier to use. While we cannot put a concrete number on how this would affect energy consumption, this move would likely increase the need for more computing power and prolong the time for which these systems are in use. It is also important to note that video is not the only expansion on the horizon: animation, character bots, 3D gaming, and Virtual Reality (VR) environments are also being worked on in the generative AI space. For instance, StabilityAI recently partnered with KrikeyAI to develop text-to-animation tools (PRNewswire 2023; StabilityAI 2023).

## Digital Waste

Now, we would like to take a step back and shift our conversation to the topic of digital waste and how these AI art systems have the potential to contribute to this type of waste. *Digital waste* (also referred to as *data waste* (Bietti and Vatanparast 2020)) is defined as "the carbon emissions, natural resource extraction, production of waste, and other harmful environmental impacts directly or indirectly attributable to data-driven infrastructures" (Bietti and Vatanparast 2020 p.2). It is our position that the large-scale adoption of generative AI art tools may contribute to the replication of our modern society's overconsumption habits of natural resources within the digital space. What we are particularly referring to is the overconsumption of the generative tools themselves, and thereby producing large amounts of data that is not only energy-intensive to generate, but also subsequently needs to be stored and maintained in data centers.

To illustrate our concerns, we collected data on how users of generative AI tools interact with these systems and utilize the produced output. We used two data sources for this purpose: a series of polls that were posted by David Holz, founder of Midjourney, posted on the main Midjourney server on various days of January 2023, as well as the previously mentioned survey which we created (Note: the polls that were posted on the Midjourney server never close, so data from these polls might change in the future. The numbers we report were accessed on February 23rd, 2023). When analyzing our data, we identified two areas of interest which we have selected for further discussion: the motivation for using these tools and the utilization of the tool and output.

## Utilization

Our survey has shown that the majority of users (57.14%) use a paid cloud service such as Midjourney, while 40.48% run a generative Art system on their home computers (the remainder of respondents use free cloud services). Most frequently, respondents claimed to create over 1000 images per week, with only 19.05% of respondents generating less than 100 images. We also identified power-users, who produce significantly larger image outputs, with one of our respondents explaining in the open-ended section of the survey that they "make 5000+ per night". The open-ended questions also showed that many users have automated the prompt-generation and use scripts to run the tools autonomously and continuously for hours. Finally, our data also showed that users frequently iterate on a single idea/prompt in order to get what they characterize as a successful piece. Approximately half of our respondents indicated that they require over 50 iterations on an idea to achieve a satisfying result, with 1 of our respondents regularly requiring over 500 interactions on a single idea.

## Motivation

Participants in our survey as well as the official polls on Midjourney's server, show reliably that most users of these tools mainly create output for themselves. Our survey indicated that 38.1% of respondents use the tools solely for themselves as entertainment with occasionally sharing creations online. While a further 21.43% of our respondents indicated that their main motivation is related to sharing their creations with others. This number varies from the data shown in Midjourney's official polls where 98% of respondents ($N = 568$) indicated that they never shared any of their creations with others and are only creating them for themselves. We hypothesize that this difference could be attributed to our sampling strategy. While the Midjourney polls reached all users of the service, our survey was targeting users that were actively engaged in AI art focused social media communities. Both data sources also show that the proportion of professional users remains a minority, with our survey indicating that only 14.29% of respondents are professional users while the Midjourney polls ($N = 3,203$) show that 35.28% of respondents had used their generations within the context of their profession.

## Digital Overconsumption

The data we have collected provides preliminary insight into how generative art systems are being deployed by users. We want to draw our reader's attention to the large amounts of data that are being generated and stored, never to be shared and consumed, and without an obvious utility. We want to lean on work by Brown & Cameron (2000), which categorizes overconsumption as a form of the common pool source dilemma, 1) where the size of the resource pool is not known, 2) access to the resources in not equally distributed among individuals and 3) individuals must make decision on their consumption of goods and services without a full picture of the quantities and types of resources required in the process. We believe that this dilemma applies here as well, as a significant portion of users are likely unaware of the total global energy consumption (and the associated GHG emissions) involved in running these generative art systems and are therefore unable to make informed decisions regarding their behavior. A single person running Stable Diffusion on their home computer might not significantly impact energy consumption, but at the scale at which these systems are currently being utilized, the impact is magnified substantially. This approach also allows us to frame this problem as an issue of education and awareness, without pointing fingers. Ideally, future cross-disciplinary approaches can lead to solutions on how to raise awareness among the different stakeholders.

# Next Steps for Stakeholders

In this section, we would like to briefly outline some possible next steps for different stakeholders who are involved in cAI.

## Users

As of this moment, most end users of these systems appear to be "everyday" users, rather than professional users who apply these systems in larger commercial settings. While we wager that this is likely to change in the future, we still think it is worth engaging these current users in the larger discourse on sustainable practices regarding digital technologies. The issues surrounding digital overconsumption and digital waste, while evident in the current usage trends of generative AI art systems, go far beyond this single technology. If we want to foster long term sustainable tech practices among users, sustained efforts in outreach and education will be required. While it is unrealistic to expect radical and rapid change in user behavior even with targeted education campaigns, we do believe that user awareness and participation in the discourse on the environmental impact of technology has the potential to create pressure on developers and encourage the creation of more sustainable systems.

## Developers

Developers of generative AI art systems, in industry and academia alike, must seriously consider the computational costs of their work, not only from a monetary perspective associated during training but also in relation to its environmental impact during inference. Discussions on Green AI have been part of the current discourse for several years at this point. Green AI, as defined by Schwartz, Dodge, Smith and Etzioni (2019), refers to improvements in AI that occur without an increase in required computational costs. While an in-depth discussion of Green AI, as well as greenwashing within the ICT (Information and Communications Technology) sector, is not within the scope of this paper, we do want to take the time to advocate for increased transparency, particularly from larger corporate entities that are now making the mass adoption of generative AI art systems possible. It is important to acknowledge that some efforts are already being made here: the environmental impact of Stable Diffusion models during training is already being estimated using the Machine Learning Impact Calculator (Lacoste et al. 2019) and made available to the public (HuggingFace n.d.). However, in order to create a clearer picture of the true environmental impact, more data is needed to accurately estimate energy consumption and carbon emissions, particularly during inference. Some recent reports have demonstrated that across the ICT sector, carbon footprint estimates have been significantly under-reporting true emission levels (Freitag, Berners-Lee, Widdicks, Knowles, Blair and Friday 2020), highlighting the need for more reliable and transparent data.

## Research Community

While the focus of our paper has been on the recent rise of generative AI art systems, we are hardly the first to attempt to raise awareness on the climate implications of AI (and the ICT sector more broadly). Instead of reiterating best practices around the development of more sustainable (generative) AI systems (for examples see: Lacoste et al. 2019; Schwartz et al. 2019; Luccioni and Hernandez-Garcia 2023), we would like to commend current on-going research efforts and encourage the creation of more spaces where these important conversations can take place. There have already been a number of conferences this year alone (such as ICCC, AIES and CVPR) that offer either special tracks and/or workshops for research on (generative) AI and our climate. With the rapid current technological developments and the rise in consumer-facing generative systems it is more important than ever that the impacts of these systems, both environmental and social, are further researched and discussed.

## Future Work

We cannot stress enough that a lot more work needs to be done in this area and that this position paper mainly serves as a tool to engage the larger cAI-community (researchers, industry, and users) in the discussion surrounding sustainable AI and how our tools impact the environment we live in. We would like to invite researchers to consider the following areas as potential new areas of inquiry:

### Expansion of Research into Climate Impacts

First and foremost, there is a need for a more comprehensive approach to the calculation of energy consumption and GHG emissions that are associated with generative art systems.

Unfortunately, we encountered frequent problems when trying to gather data to support our thoughts in this paper. There is only a small amount of publicly available data on the use of these systems and increased transparency is required for researchers to estimate the climate implications more accurately. The following data would be required for future work: 1) daily active users for each system, 2) total daily image output for each system, 3) hardware used to process data (GPUs, cloud services), 4) location of users (as the GHG emissions associated with electricity generation varies by country and energy source (Luccioni and Hernandez-Garcia 2023)) and 5) any initiatives that might have been taken by OpenAI, Midjourney and StabilityAI to reduce climate implications.

### Human Interaction with Generative AI Systems

Future work should look further into the behavior patterns and motivations of users of these generative art systems.

For instance, we hypothesize that one possible explanation for the recent explosion of use, is related to uses and gratification theory (Katz, Blumer and Gurevitch 1973), a widely cited framework to study how media has the ability to satisfy a person's needs and desires, which leads to the continued and prolonged consumption. It has already been used to explain the addictive nature of social media platforms like TikTok (Montag, Yang and Elhai 2021). Particularly, a need for escapism has been linked to increased consumption of digital content (Omar and Dequan 2020). It should be investigated whether these ideas also apply in the context of creative AI tools which allow almost instant content creation and consumption, which would not be possible otherwise. These tools allow users to create artworks at a quality that would have required high levels of fine arts skills and a significant time investment (both to acquire the skills and then to execute the artwork). Further study of how these tools interact with our cognitive reward system also demands heightened attention as these tools move from mainly still imagery to full video creation and eventually 3D VR environments. We have conjectured that there could be potential dangers here (as well as opportunities for social good), related to the notion of escapism and the ability of users to now create artificial worlds that serve as a virtual sanctuary from reality.

While this is an issue for further studies – this general hypothesis around these systems having the potential of satisfying essential human needs such as personal/creative expression and connectedness, does speak to their explosive growth and the possible issue of encountering increased difficulty and resistance in trying to educate users on the negative environmental implications of their actions.

Due to the widespread adoption of these tools, such work could also provide valuable insights into everyday human-AI interaction and meaning making in the digital space. This could further our understanding of the societal impact of these new technologies.

## Limitations

We made our best efforts to find up-to-date data to back up our analysis and calculations. However, it is important to acknowledge that it is currently difficult to obtain a complete picture of the demonstrated problems due to a lack of availability of reliable data. In the Future Work section, we outlined the kinds of data that would be required to get more precise estimates on the climate implications. It is also important to discuss the possibility that the current numbers that we presented in this work merely reflect a snapshot of user behavior during a time of immense "hype" around these systems due to their novelty. However, while it is most likely correct to assume that the current level of fascination that many users have with these systems will die down eventually, these systems are likely to be adopted within many professional creative contexts in the future and while there might be a demographic shift in the users base and the types of application we encounter, we predict that the overall use of generative AI art systems will likely increase over the month and years to come.